\documentclass[aps,prl,twocolumn,superscriptaddress]{revtex4}
\usepackage{graphicx}
\usepackage{amssymb}
\usepackage{amsmath}
\usepackage{graphicx}
\usepackage{epsfig}
\usepackage{array}
\usepackage{color}
\usepackage{multirow}

\begin{document}

\title{Quasi-two-dimensional Fermi surfaces and unitary spin-triplet pairing \\
in the heavy fermion superconductor UTe$_2$}

\author{Yuanji Xu}
\author{Yutao Sheng}
\affiliation{Beijing National Laboratory for Condensed Matter Physics and
Institute of Physics, Chinese Academy of Sciences, Beijing 100190, China}
\affiliation{University of Chinese Academy of Sciences, Beijing 100049, China}
\author{Yi-feng Yang}
\email[]{yifeng@iphy.ac.cn}
\affiliation{Beijing National Laboratory for Condensed Matter Physics and
Institute of Physics, Chinese Academy of Sciences, Beijing 100190, China}
\affiliation{University of Chinese Academy of Sciences, Beijing 100049, China}
\affiliation{Songshan Lake Materials Laboratory, Dongguan, Guangdong 523808, China}

\date{\today}

\begin{abstract}
We report first-principles and strongly-correlated calculations of the newly-discovered heavy fermion superconductor UTe$_2$. Our analyses reveal three key aspects of its magnetic, electronic, and superconducting properties, that include: (1) a two-leg ladder-type structure with strong magnetic frustrations, which might explain the absence of long-range orders and the observed magnetic and transport anisotropy; (2) quasi-two-dimensional Fermi surfaces composed of two separate electron and hole cylinders with similar nesting properties as in UGe$_2$, which may potentially promote magnetic fluctuations and help to enhance the spin-triplet pairing; (3) a unitary spin-triplet pairing state of strong spin-orbit coupling at zero field, with point nodes presumably on the heavier hole Fermi surface along the $k_x$-direction, in contrast to the previous belief of non-unitary pairing. Our proposed scenario is in excellent agreement with latest thermal conductivity measurement and provides a basis for understanding the peculiar magnetic and superconducting properties of UTe$_2$.
\end{abstract}

\maketitle

Recent discovery of superconductivity in UTe$_2$ with $T_c=1.6\,$K at ambient pressure and zero magnetic field has attracted intensive interest in the heavy fermion community \cite{Ran2019,Aoki2019}. Muon spin relaxation/rotation ($\mu$SR) experiments revealed strong ferromagnetic fluctuations coexisting with the superconductivity \cite{Sundar2019}. A large upper critical field was found to exceed the Pauli paramagnetic limit and resemble that in UGe$_2$, UCoGe, and URhGe \cite{Sheikin2001,Aoki2009,Levy2005,AokiIshida2019}. But different from these latter compounds \cite{Saxena2000,Aoki2001,Huy2007}, superconductivity in UTe$_2$ emerges out of a paramagnetic normal state. It was hence proposed to be at the verge of a ferromagnetic phase and have exotic non-unitary spin-triplet pairing that breaks the time-reversal symmetry  \cite{Ran2019,Aoki2019}. The nuclear magnetic resonance (NMR) Knight shift indeed remains constant below $T_c$ and supports the spin-triplet pairing \cite{Ran2019}. Further analysis of the specific heat ($\sim T^3$), thermal conductivity ($\sim T^3$), and penetration depth ($\sim T^2$) has led to the proposal of point nodes in superconducting gap \cite{Metz2019}. On the other hand, a large extrapolated value for the residual Sommerfeld coefficient ($\gamma_0 = 55\,$mJ/mol K$^2$) seems to indicate that only half of the electrons are gapped \cite{Ran2019}. By contrast, thermal conductivity revealed a vanishingly small fermionic carrier density at zero temperature limit \cite{Metz2019}. Upon applying the magnetic field, two field-reentrant superconducting phases emerge, possibly associated with some field-driven metamagnetic transition or Fermi surface instability \cite{Knafo2019,Miyake2019,Knebel2019,SRan2019,Niu2019,Imajo2019}. It was even proposed that the system might host topological excitations \cite{Ran2019,Kozii2016,Sau2012}, making it a rich playground for exploring exotic heavy fermion phenomena.

In contrast to the rapid progress in superconducting measurements, the magnetic and electronic structures of UTe$_2$ remain unclear in theory. Previous band calculations predicted a semiconducting normal state, in contradiction with the observed metallicity in transport measurements \cite{Aoki2019}. It is evident that Fermi surface topology is crucial for superconducting pairing and its nodal properties \cite{Bastien2016,Yu2018}. In particular, one may wonder if the proposed non-unitary spin-triplet pairing is indeed capable of explaining the observed point nodes in experiment.

In this work, we report first-principles and strongly-correlated electronic structure calculations for UTe$_2$ using the density functional theory (DFT) \cite{Blaha2014,Perdew1996} and dynamical mean-field theory (DMFT) \cite{Kotliar2006,Georges1996,Haule2010,Held2008} approaches. We show that including both the Coulomb interaction and the spin-orbit coupling (SOC) can suppress the semiconducting gap and produce flat $f$ electron/hole bands across the Fermi energy. Our analyses reveal three key aspects of the magnetic, electronic, and superconducting  properties of UTe$_2$. Magnetic calculations find a two-leg ladder-type structure with strong magnetic frustrations, which might be responsible for the absence of long-range orders and the observed magnetic and transport anisotropy. The calculated Fermi surfaces are of quasi-two-dimensional (2D) character and contain two separate electron and hole cylinders with nesting properties similar to UGe$_2$ that may potentially promote magnetic fluctuations and enhance the spin-triplet pairing. Simple group theoretical analysis excludes previous proposal of non-unitary pairing and suggests a unitary spin-triplet pairing state of strong-SOC representation with point nodes presumably on the heavier hole Fermi surface along the $k_x$-direction. Our results are in excellent agreement with latest thermal conductivity measurement and provide a promising basis for understanding the key physics of UTe$_2$.

\begin{figure}[b]
\begin{center}
\includegraphics[width=0.48\textwidth]{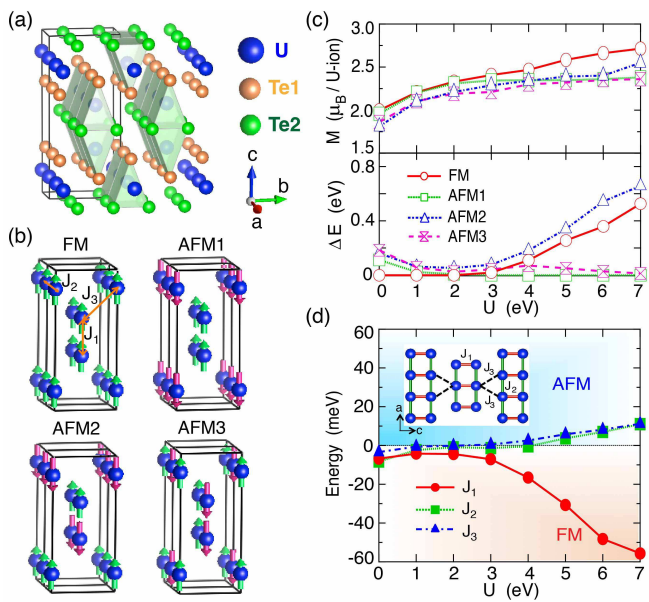}
\caption{(a) Illustration of the crystal structure of UTe$_2$, showing the two-leg U-ladders surrounded by face-shared Te-prisms. (b) Four chosen magnetic configurations in a $2\times1\times1$ supercell for calculations of the exchange couplings $J_i$ between U-ions up to the 3rd nearest neighbors. (c) The calculated magnetic moment of U-ion and the energy differences (per supercell) relative to the lowest energy state as a function of $U$. (d) The derived values of $J_i$ with varying $U$, showing a dominant ferromagnetic (FM) rung coupling $J_1$ and much smaller antiferromagnetic (AFM) couplings $J_2$ on the leg and $J_3$ between ladders at large $U$. The inset illustrates how  magnetic frustrations are induced between ladders by their relative shift of half lattice constant 0.5$a$ along the $a$-axis.}
\label{fig1}
\end{center}
\end{figure}

We first focus on structural and magnetic properties of UTe$_2$. Different from tellurium-deficient UTe$_{2-x}$ \cite{Haneveld1970,Ellert1971,Stowe1996,Haneveld1969,Stowe1997}, the stoichiometric UTe$_2$ adopts an orthorhombic structure with the space group $Immm$ and the lattice parameters, $a=4.16\,$\AA{}, $b=6.12\,$\AA{}, and $c=13.96\,$\AA{} \cite{Ikeda2006}. Each U-ion is surrounded by six Te-ions, forming together a trigonal prism. As illustrated in Fig.~\ref{fig1}(a), the two U-chains form a two-leg ladder along the $a$-axis, enclosed by the face-shared prisms. The rung distance is about 3.78 \AA{} and smaller than the U distance of $4.16\,$\AA{} on the leg. The shortest U distance between two ladders is farther away and about 4.89 \AA{}. Thus the two-leg ladders may be viewed as the basic building block of the U-lattice. To get an idea about the magnetic interactions of this ladder system, we calculated the energies of four chosen magnetic configurations in Fig.~\ref{fig1}(b) and subtracted the exchange couplings, $J_i$, up to the 3rd nearest neighbors. Figure~\ref{fig1}(c) plots the calculated magnetic moment of U-ion and the energy differences relative to the lowest energy state with varying Coulomb interaction. Contrary to the usual expectation, among all four configurations, FM has the lowest energy only at small $U$. For large $U$, AFM1 and AFM3 approach the same energy, indicating the presence of magnetic frustrations to be discussed in more detail below. For all configurations, the moment is close to saturation at large $U$ and reveals somewhat over two polarized $f$ electrons per U-ion, consistent with its expected valence \cite{Stowe1997}. The values of $J_i$ can then be estimated by fitting the magnetic energies with the effective Hamiltonian, $H= \sum_{\langle lm\rangle}J_{lm}{\bf S}_{l} \cdot {\bf S}_{m} $, where ${\bf S}_{l/m}$ are the polarized spins and the magnetocrystalline anisotropy was neglected for simplicity.

Figure~\ref{fig1}(d) plots the estimated values of $J_i$ as a function of $U$. For large $U$ ($\ge 6\,$eV), which is typical for $f$ electrons \cite{Shim2009,Yin2011}, we find a dominant ferromagnetic rung coupling $J_1$ compared to the much smaller $J_2$ on the leg and $J_3$ between the ladders. The antiferromagnetic nature of $J_2$ and $J_3$ seems to be supported by the negative Weiss temperature ($78-126\,$K) derived from the Curie-Weiss fit of the magnetic susceptibilities along all three directions \cite{Ikeda2006}. The fact that they all have the same order of magnitude of about $5-10\,$meV supports our choice of a large $U$ and also suggests that the ladder structure might actually be responsible for the anisotropy observed in magnetic and transport properties \cite{Ran2019,Ikeda2006,Tokunaga2019}. Moreover, as shown in the inset of Fig.~\ref{fig1}(d), for comparable and antiferromagnetic $J_2$ and $J_3$,  the relative shift of half lattice constant (0.5$a$) along the $a$-axis induces frustrated interactions between ladders. As a consequence, AFM1 and AFM3 become almost degenerate at $U=7\,$eV. The magnetic frustration and reduced dimensionality of the ladders might be a potential origin for the suppressed magnetic orders and observed metamagnetic transitions in UTe$_2$. For a moderate $U$ of about $4\,$eV, both $J_2$ and $J_3$ become negligible, and the ladders are disassembled into a gas of ``ferromagnetic pairs'', inconsistent with experiment. In either case, the situation seems very different from UGe$_2$, UCoGe, URhGe, or other layered superconductors. The two-leg ladder structure has been extensively studied in cuprate superconductors \cite{Dagotto1994,Dagotto1999} and lately also found in some Fe-based superconductors \cite{Takahashi2015,Chi2016,Zhang2017}. It has attracted much interest over the past years in both theory and experiment as an alternative and simpler platform for unconventional superconductivity. We anticipate that the frustrated ladder structure also plays a key role for the peculiar magnetism and superconductivity in UTe$_2$ and expect rich magnetic ground states tuned by external field or pressure.

\begin{figure}
\begin{center}
\includegraphics[width=0.48\textwidth]{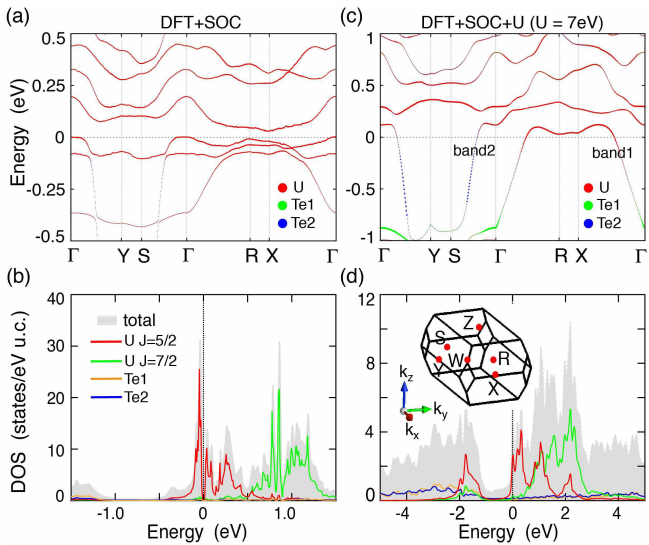}
\caption{Comparison of the calculated band structures and density of states with DFT+$U$ for (a,b) $U=0$ and (c,d) $U=7\,$eV. The colors represent the contributions from different U or Te orbitals. We see a small semiconducting gap of about 10 meV for $U=0$ and flat metallic bands crossing the Fermi level for $U=7\,$eV. In both cases, the total density of states near the Fermi energy is dominated by the $J=5/2$ manifold of the U-5$f$ orbitals. The $J=7/2$ manifold is pushed to higher energies by about $1.0-1.5\,$eV. The inset shows the high symmetry points in the first Brillouin zone.}
\label{fig2}
\end{center}
\end{figure}

We now proceed to discuss the electronic band structures of UTe$_2$. Previous DFT calculations predicted a semiconducting ground state for the paramagnetic phase \cite{Aoki2019}. This is reproduced in Figs.~\ref{fig2}(a) and \ref{fig2}(b) with $U=0$ but contradicts the experimental observation of metallicity. We find that by including both the Coulomb interaction and SOC, the band gap can be closed and the ground state can be tuned into a metal. As shown in Figs.~\ref{fig2}(c) and \ref{fig2}(d) for $U=7\,$eV, two flat metallic bands of dominant $f$ character now cross the Fermi level. Accordingly, a sharp peak appears near the Fermi energy in the total density of states. There are two types of charge carriers here. The $\Gamma$-$R$-$X$-$\Gamma$ path is mainly along the $k_x$-direction in the Brillouin zone and gives the hole band (denoted as band1), while the $\Gamma$-$Y$-$S$-$\Gamma$ path presents the electron band (band2) dispersed along the $k_y$-direction. The two bands belong to the $J=5/2$ manifold of U 5$f$ electrons and originate from the hybridization with two inequivalent Te-ions, respectively. The $J=7/2$ manifold is located at much higher energy with the spin-orbit splitting of about 1.5 eV \cite{Halilov1991}. Needless to say,  electronic correlations are essential for the metallicity of UTe$_2$.

\begin{figure}
\begin{center}
\includegraphics[width=0.49\textwidth]{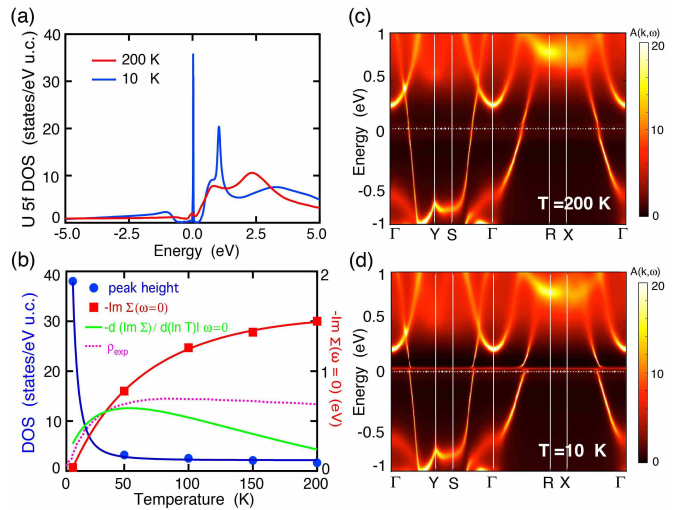}
\caption{(a) The calculated density of states of U 5$f$ electrons with self-consistent DFT+DMFT, showing a sharp quasiparticle peak at 10 K that is suppressed at 200 K. (b) Temperature evolution of the peak height and the imaginary part of the self-energy at the Fermi energy. The temperature derivative of the latter is compared with the measured resistivity \cite{Shlyk1999}, showing similar tendency below the coherence temperature of about $50\,$K. (c) and (d) Comparison of the spectral functions at 200 K and 10 K. Extremely flat heavy electron/hole bands are seen to emerge in a narrow window around the Fermi energy at low temperature. The background colors reflect the intensity of the total spectral function.}
\label{fig3}
\end{center}
\end{figure}

\begin{figure}
\begin{center}
\includegraphics[width=0.48\textwidth]{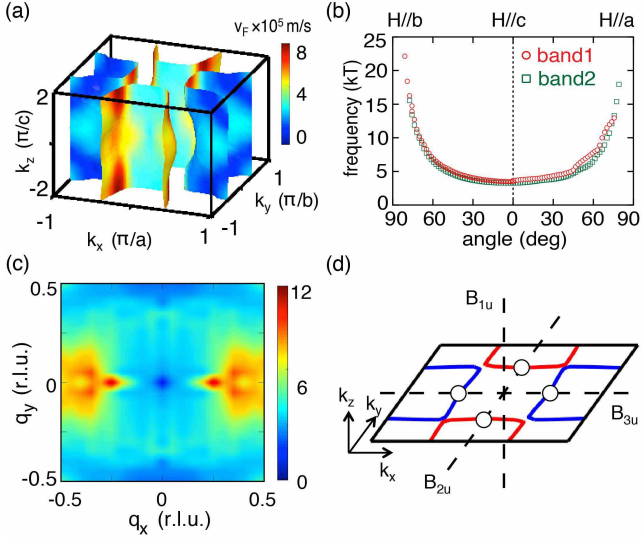}
\caption{(a) The calculated Fermi surfaces with two separate quasi-2D electron and (heavier) hole cylinders; (b) The predicted quantum oscillation frequencies with field rotating from the $c$-axis to $a$ or $b$-axes for dHvA measurements; (c) Real part of the dynamical susceptibility at zero frequency limit, showing nesting properties along the $k_x$-direction near half of the reciprocal lattice unit (r.l.u.). All data were based on DFT+$U$ calculations as in Fig.~\ref{fig2}(c) but with 8000 $\bf k$-points in order to get high-quality plot. The color bars represent the value of the Fermi velocity in (a) and the magnitude (arbitrary unit) of the susceptibility in (c). (d) Illustration of the nodal properties of the candidate strong-SOC pairing states, showing point nodes for $B_{2u}$ and $B_{3u}$ representations on the calculated electron and hole Fermi surfaces, respectively. The gap magnitude is zero on the dashed lines given by $k_x=k_y=0$ for $B_{1u}$, $k_x=k_z=0$ for $B_{2u}$, and $k_y=k_z=0$ for $B_{3u}$.}
\label{fig4}
\end{center}
\end{figure}

The above electronic structures are further confirmed by our DFT+DMFT calculations \cite{Kotliar2006,Georges1996,Haule2010,Held2008}. We  used the continuous-time quantum Monte Carlo method (CTQMC) as the impurity solver and took the nominal double counting for the full charge self-consistent calculations \cite{Werner2006,Haule2007}. The real-frequency self-energy was obtained by analytic continuation. Figure~\ref{fig3}(a) plots the density of states of U-5$f$ electrons for $U=8\,$eV and $J=0.6\,$eV following previous calculations for uranium oxides \cite{Shim2009,Yin2011}. A large $U$ is typically needed here because of the Coulomb screening effect in summing over all local diagrams, but our qualitative results are unchanged with its variation in a reasonable range. Anyway, we see as expected a sharp quasiparticle peak developing near the Fermi energy at 10 K, which is suppressed at 200 K. For experimental comparison, we also plot the imaginary part of the self-energy, whose temperature derivative at the Fermi energy ($\omega=0$) resembles that of the quasiparticle scattering rate \cite{Shim2007}. Indeed, it follows roughly the measured resistivity at low temperatures with the correct coherence temperature of about 50 K \cite{Ran2019,Aoki2019,Shlyk1999}. The height of the quasiparticle peak is also plotted in Fig.~\ref{fig3}(b) and seen to increase substantially below the same temperature, implying the rapid development of heavy electron states once the coherence sets in \cite{Yang2008}. Figures~\ref{fig3}(c) and \ref{fig3}(d) compare the spectra at 200 K and 10 K. While the $f$ electrons are well localized at high temperatures, we see extremely flat bands emerge near the Fermi energy at 10 K and hybridize with the conduction bands. The overall features are consistent with DFT+$U$ calculations, except that the bands are more strongly renormalized. A tentative fit gives the magnitude of about 25 meV for the electron band and 120 meV for the hole one to be examined in optical measurement. Typically, the larger gap size of the hole band means a heavier quasiparticle effective mass of the hole carriers \cite{Dordevic2001,Chen2016}.

For further analysis of the superconductivity, we plot in Fig.~\ref{fig4}(a) the calculated Fermi surfaces for UTe$_2$. Interestingly, we see two slightly corrugated cylinders that are only weakly dispersive along the $k_z$-direction so that the whole Fermi surfaces are essentially quasi-2D. This was initially not expected and may be ascribed to the strong rung coupling of U-ladders and the layer structures of surrounding Te-ions. The two cylindrical Fermi surfaces originate from the flat electron and hole bands, respectively. Once again, we see the holes have a smaller Fermi velocity and thus a heavier quasiparticle effective mass. For experimental examination, we also present in Fig.~\ref{fig4}(b) the de Haas-van Alphen (dHvA) quantum oscillation frequencies for field rotating from $c$-axis to $a$ or $b$-axes \cite{Rourke2012}. Its monotonic increase away from the $c$-axis and divergence along the perpendicular axes demonstrate the quasi-2D character of the Fermi surfaces that could be easily verified in future measurements. Figure~\ref{fig4}(c) also plots the real part of the dynamical susceptibility derived under the random phase approximation (RPA). The maxima imply the Fermi surface nesting along the $k_x$-direction, similar to that found in UGe$_2$ \cite{Shick2001}. It has been argued that this could potentially promote magnetic fluctuations and help to enhance the spin-triplet pairing \cite{Watanabe2002,Kang2018}. The reason that the $k_y$-direction is less nested is probably associated with the zig-zag atomic structure along the crystalline $b$-axis.

The Fermi surface topology provides a primary basis for discussing the superconducting pairing symmetry. In analogy with UGe$_2$, UCoGe, and URhGe, previous experimental analysis has suggested that UTe$_2$ might have a non-unitary spin-triplet pairing state with point nodes \cite{Ran2019}. This immediately led to the proposal of a weak-SOC pairing state, ${\bf d}({\bf k})=\varphi({\bf k})(1,i,0)$ \cite{Aoki2019}, which is an extreme ``equal spin pairing" state of half-gapped superconductivity and has been well studied for the A1-phase of $^3$He superfluid. The strong-SOC pairing states were all excluded because the time-reversal symmetry can only be broken in a multidimensional representation, which is forbidden in the point group $D_{2h}$ \cite{Yip1993}. For clarity, we list in Table~\ref{tab1} all odd-parity representations of the point group $D_{2h}$ and their nodal structure on our calculated Fermi surfaces. Regardless of the unitary property of the ${\bf d_0}$-vector, all four weak-SOC representations predict line nodes, which disagrees with the experimental implication of point nodes. This assertion does not depend on any details other than the quasi-two-dimensionality of the Fermi surfaces.

\begin{table}[t]
\caption{\label{tab1} The odd-parity pairing states for all irreducible representations of the point group $D_{2h}$. For weak SOC, the ${\bf d}$-vector has the form, ${\bf d}({\bf k})=\varphi({\bf k}){\bf d_0}$, where $\varphi({\bf k})$ is the basis fucntion and ${\bf d_0}$ is a constant vector. The pairing state is called unitary if ${\bf d}_0\times{\bf d}_0=0$ and non-unitary otherwise. For strong SOC, all representations are one dimensional and therefore unitary. The nodal properties are obtained by projecting the basis functions on the calculated Fermi surfaces. $\eta_i$ are constant prefactors and ${\bf \hat x}$, ${\bf \hat y}$, ${\bf \hat z}$ are unit vectors along three axes.\\}
\centering
\begin{tabular}{|c|c|c|c|}
\hline
SOC	& \ \ \ reps \ \ \ 	& ${\bf d}$-vector 				&\ \ \  node\ \ \ 		\\
\hline
\multirow{4}{*}{weak}	    & $A_{u}$			& $k_xk_yk_z{\bf d}_0$	& lines		\\
        & $B_{1u}$			& $k_z{\bf d}_0$			& lines		\\
	& $B_{2u}$			& $k_y{\bf d}_0$			& lines		\\
	& $B_{3u}$			& $k_x{\bf d}_0$			& lines		\\
\hline
\multirow{4}{*}{strong} & $A_{u}$	& $\eta_1 k_x{\bf \hat x} + \eta_2 k_y{\bf \hat y} + \eta_3 k_z{\bf \hat z}$			& none	\\
	& $B_{1u}$	& $\eta_1 k_y{\bf \hat x} + \eta_2 k_x{\bf \hat y} + \eta_3 k_xk_yk_z{\bf \hat z}$	& none	\\
	& $B_{2u}$	& $\eta_1 k_z{\bf \hat x} + \eta_2 k_xk_yk_z{\bf \hat y} + \eta_3 k_x{\bf \hat z}$	& points	\\
	& $B_{3u}$	&\ \ \  $\eta_1 k_xk_yk_z{\bf \hat x} + \eta_2 k_z{\bf \hat y} + \eta_3 k_y{\bf \hat z}$	\ \ \ & points	\\
\hline
\end{tabular}
\end{table}

To solve this dilemma, we point out that superconductivity in UTe$_2$ is different from that in UGe$_2$ and actually born out of a paramagnetic normal state that does not necessarily break the time-reversal symmetry \cite{Sundar2019}. If we are allowed to release the requirement of non-unitarity, we can see that among all four representations of strong SOC, two ($A_u$ and $B_{1u}$) will be fully gapped on our Fermi surfaces, and only $B_{2u}$ and $B_{3u}$ representations can have point nodes. Thus our calculated Fermi surfaces demand a unitary spin-triplet pairing state of either $B_{2u}$ or $B_{3u}$ representation. As illustrated in Fig.~\ref{fig4}(d), this will give point nodes on one of the two Fermi surfaces, presumably the heavier and more nested hole Fermi surface (the $B_{3u}$ representation). This is in excellent agreement with the observed anisotropy in thermal conductivity, which also suggests point nodes along the $a$-axis \cite{Metz2019}. Moreover, since this is no longer ``equal spin pairing", there would be no half-gapped Fermi surface unless one of the two cylindrical Fermi surfaces does not participate in the superconductivity. This is, however, usually very unlikely in reality. Actually, latest measurements of thermal conductivity did indeed suggest vanishingly small residual fermionic carriers at zero temperature limit. The specific heat was found to exhibit a logarithmic upturn below 300 mK, and the large residual Sommerfeld coefficient was attributed to potentially localized or strongly scattered divergent quantum critical contributions \cite{Metz2019}. This is a supportive evidence for our scenario, although the exact source for the divergence awaits further experimental elaboration.

To summarize, we have performed first-principles and strongly-correlated calculations for the electronic and magnetic properties of the newly-discovered heavy fermion superconductor UTe$_2$. We find that electronic correlations are essential in order to explain its metallicity. Further analyses reveal three key aspects of its magnetic, electronic, and superconducting properties. These include a ladder-type structure with strong magnetic  frustrations and quasi-2D Fermi surfaces composed of two separate electron and hole cylinders, which are nested along the $k_x$-direction similar to UGe$_2$ and might potentially enhance magnetic fluctuations and the spin-triplet pairing. The quasi-two-dimensionality puts strict constraint on the candidate pairing state. We argue that previously proposed non-unitary pairing is inconsistent with the experimental implication of point nodes and therefore excluded. Instead, we propose a unitary spin-triplet pairing state of strong SOC, with point nodes presumably on the heavier hole Fermi surface along the $k_x$-direction. This scenario is in excellent agreement with latest thermal conductivity measurement and may therefore provide a useful basis for understanding the peculiar magnetism and superconductivity in UTe$_2$.

This work was supported by the National Natural Science Foundation of China (NSFC Grant No. 11774401, No. 11974397), the National Key R\&D Program of China (Grant No. 2017YFA0303103), the State Key Development Program for Basic Research of China (Grant No. 2015CB921303), the National Youth Top-notch Talent Support Program of China, and the Youth Innovation Promotion Association of CAS.

\end{document}